\documentclass[superscriptaddress,onecolumn,secnumarabic,
amssymb,amsmath,nobibnotes,aps,prd,showkeys,showpacs,nofootinbib]{revtex4}

\usepackage[latin1]{inputenc}
\usepackage{graphicx}
\usepackage[english]{babel}

\usepackage{amsmath}
\usepackage{amssymb}
\usepackage{amsfonts}
\usepackage{colordvi}
\usepackage{psfrag}
\usepackage{color}

\allowdisplaybreaks[4]

\begin{document}

\newcommand\cL{\mathcal{L}}
\newcommand\be{\begin{equation}}
\newcommand\ee{\end{equation}}
\newcommand\bea{\begin{eqnarray}}
\newcommand\eea{\end{eqnarray}}
\newcommand\beq{\begin{eqnarray}}
\newcommand\eeq{\end{eqnarray}}
\newcommand\tr{{\rm tr}\, }
\newcommand\nn{\nonumber \\}
\newcommand\e{{\rm e}}

\newcommand\bef{\begin{figure}}
\newcommand\eef{\end{figure}}
\newcommand{\ans}{ansatz }
\newcommand{\eeqn}{\end{eqnarray}}
\newcommand{\bd}{\begin{displaymath}}
\newcommand{\ed}{\end{displaymath}}
\newcommand{\mat}[4]{\left(\begin{array}{cc}{#1}&{#2}\\{#3}&{#4}
\end{array}\right)}
\newcommand{\matr}[9]{\left(\begin{array}{ccc}{#1}&{#2}&{#3}\\
{#4}&{#5}&{#6}\\{#7}&{#8}&{#9}\end{array}\right)}
\newcommand{\matrr}[6]{\left(\begin{array}{cc}{#1}&{#2}\\
{#3}&{#4}\\{#5}&{#6}\end{array}\right)}
\newcommand{\cvb}[3]{#1^{#2}_{#3}}
\newcommand\lsim{\raise0.3ex\hbox{$\;<$\kern-0.75em\raise-1.1ex
e\hbox{$\sim\;$}}}
\newcommand\gsim{\raise0.3ex\hbox{$\;>$\kern-0.75em\raise-1.1ex
\hbox{$\sim\;$}}}
\def\abs#1{\left| #1\right|}
\newcommand\simlt{\mathrel{\lower2.5pt\vbox{\lineskip=0pt\baselineskip=0pt
           \hbox{$<$}\hbox{$\sim$}}}}
\newcommand\simgt{\mathrel{\lower2.5pt\vbox{\lineskip=0pt\baselineskip=0pt
           \hbox{$>$}\hbox{$\sim$}}}}
\newcommand\unity{{\hbox{1\kern-.8mm l}}}
\newcommand{\eps}{\varepsilon}
\newcommand\ep{\epsilon}
\newcommand\ga{\gamma}
\newcommand\Ga{\Gamma}
\newcommand\om{\omega}
\newcommand\omp{{\omega^\prime}}
\newcommand\Om{\Omega}
\newcommand\la{\lambda}
\newcommand\La{\Lambda}
\newcommand\al{\alpha}
\newcommand{\ov}{\overline}
\renewcommand{\to}{\rightarrow}
\renewcommand{\vec}[1]{\mathbf{#1}}
\newcommand{\vect}[1]{\mbox{\boldmath$#1$}}
\newcommand\tm{{\widetilde{m}}}
\newcommand\mcirc{{\stackrel{o}{m}}}
\newcommand{\Dm}{\Delta m}
\newcommand{\dm}{\varepsilon}
\newcommand{\tanb}{\tan\beta}
\newcommand{\nbar}{\tilde{n}}
\newcommand\PM[1]{\begin{pmatrix}#1\end{pmatrix}}
\newcommand{\up}{\uparrow}
\newcommand{\down}{\downarrow}
\newcommand\omE{\omega_{\rm Ter}}
%
%%%%%%%%%%     mauri    %%%%%%%%%%%%%%%%%%%%%%%%%%%%%%%%%

\newcommand{\Dsusy}{{susy \hspace{-9.4pt} \slash}\;}
\newcommand{\DCP}{{CP \hspace{-7.4pt} \slash}\;}
\newcommand{\mc}{\mathcal}
\newcommand{\gr}{\mathbf}
\renewcommand{\to}{\rightarrow}
\newcommand{\gtc}{\mathfrak}
\newcommand{\wh}{\widehat}
\newcommand{\br}{\langle}
\newcommand{\kt}{\rangle}

%%%%%%%%%%%%%%%%%%%%%%%%%%%%%%%%%%%%%%%%%%%%%%%%%%%%%%%%%%

% barbara Ricci  %definizione di minore e maggiore simile
\def\lsim{\mathrel{\mathop  {\hbox{\lower0.5ex\hbox{$\sim$}
\kern-0.8em\lower-0.7ex\hbox{$<$}}}}}
\def\gsim{\mathrel{\mathop  {\hbox{\lower0.5ex\hbox{$\sim$}
\kern-0.8em\lower-0.7ex\hbox{$>$}}}}}
%%%%%%%%%%%%%%%%%%%%%%%%%%%%%%%%%%

\newcommand\de{\partial}
\newcommand\brf{{\mathbf f}}
\newcommand\bbf{\bar{\bf f}}
\newcommand\bF{{\bf F}}
\newcommand\bbF{\bar{\bf F}}
\newcommand\bA{{\mathbf A}}
\newcommand\bB{{\mathbf B}}
\newcommand\bG{{\mathbf G}}
\newcommand\bI{{\mathbf I}}
\newcommand\bM{{\mathbf M}}
\newcommand\bY{{\mathbf Y}}
\newcommand\bX{{\mathbf X}}
\newcommand\bS{{\mathbf S}}
\newcommand\bb{{\mathbf b}}
\newcommand\bh{{\mathbf h}}
\newcommand\bg{{\mathbf g}}
\newcommand\bla{{\mathbf \la}}
\newcommand\bmu{\mathbf m }
\newcommand\by{{\mathbf y}}
\newcommand\bsig{\mbox{\boldmath $\sigma$} }
\newcommand\bunity{{\mathbf 1}}
\newcommand\cA{\mathcal{A}}
\newcommand\cB{\mathcal{B}}
\newcommand\cC{\mathcal{C}}
\newcommand\cD{\mathcal{D}}
\newcommand\cF{\mathcal{F}}
\newcommand\cG{\mathcal{G}}
\newcommand\cH{\mathcal{H}}
\newcommand\cI{\mathcal{I}}
\newcommand\cN{\mathcal{N}}
\newcommand\cM{\mathcal{M}}
\newcommand\cO{\mathcal{O}}
\newcommand\cR{\mathcal{R}}
\newcommand\cS{\mathcal{S}}
\newcommand\cT{\mathcal{T}}
\newcommand\eV{\mathrm{eV}}

\title{Evaporation and anti-evaporation instability of a Schwarzschild-de Sitter braneworld: the case of five-dimensional F(R) gravity}

\author{Andrea Addazi}
\affiliation{ Dipartimento di Fisica, Universit\`a di L'Aquila, 67010 Coppito AQ, Italy}
\affiliation{Laboratori Nazionali del Gran Sasso (INFN), 67010 Assergi AQ, Italy,}

\author {Shin'ichi Nojiri}
\affiliation{Department of Physics, Nagoya University, Nagoya 464-8602, Japan,} 
\affiliation{Kobayashi-Maskawa Institute for the Origin of Particles and the Universe, 
Nagoya University, Nagoya 464-8602, Japan}

\author {Sergei Odintsov}
\affiliation{Instituci\'o Catalana de Recerca i Estudis Avancats (ICREA), Barcelona, Spain,} 
\affiliation{Institut de Ciencies de l'Espai (IEEC-CSIC), Campus UAB, Carrer de Can Magrans, s/n
08193 Cerdanyola del Valles, Barcelona, Spain} 

%\date{\today}
%\vspace{1cm}

\begin{abstract}
We study the problem of a four-dimensional brane lying in the five-dimensional degenerate 
Schwarzschild-de Sitter (Nariai) black hole, in five-dimensional $F(R)$-gravity.
We show that there cannot exist the brane in the Nariai bulk space except the case that the brane 
tension vanishes.
We demonstrate that the five-dimensional Nariai bulk is unstable in a large region of the parameter 
space.
In particular, the Nariai bulk can have classical (anti-)evaporation instabilities.
The bulk instability back-reacts on the four-dimensional brane, in case that the brane tension 
vanishes, and the unstable modes propagate in their world-volume.
%This opens new possilibities for the interpretation of our results in frames of dS/CFT 
%correspondence.
\end{abstract}


\maketitle

\section{Introduction}

The possibility that our Universe could be a brane lying in a higher dimensional bulk is strongly 
motivated by string theory and was largely explored in literature 
(see Refs.~\cite{Maartens:2010ar,Brax:2003fv} for reviews on these subjects).
Some observational experiments are currently searching for the manifestation of extra dimensions.
On the other hand, the study of brane in a higher dimensional de Sitter bulk is strongly motivated as 
in context of string theory (see, for instance, 
Refs.~\cite{Strominger:2001pn,Hawking:2000bb,Nojiri:2000gb,Cvetic:2001bk,
Odintsov:2002zi,Nojiri:2002qt,Addazi:2016wpz}).
In particular, the extension of the higher-dimensional theory of gravity from the Einstein-Hilbert 
formulation to ghost-free $F(R)$-gravity is highly motivated by 
$O\left(\left(\alpha'\right)^{n}\right)$ perturbative corrections to string amplitude as well as 
by possible non-perturbative effects like (Euclidean) D-brane or worldsheet instantons (see 
Ref.~\cite{Bianchi:2009ij} for a useful review on these aspects) \footnote{See also 
Refs.~\cite{Nojiri:2006ri,Nojiri:2010wj,Clifton:2011jh}
for useful reviews in extended theories of gravity.}.
In this letter, we explore the dynamics of brane lying in the higher dimensional Nariai black hole bulk, 
in the context of five-dimensional $F(R)$-gravity.

We find some surprising results;
\begin{enumerate}
\item The Nariai black hole background is unstable in a large space of parameters of the 
$F(R)$-gravity. In particular, (anti-)evaporation instabilities and evaporation/anti-evaporation 
iterative oscillations studied in four dimensional are retried in this context 
\cite{Bousso:1997wi,Addazi:2016prb}.
\item The background instabilities change the brane dynamics provoking worldsheet instabilities.
\end{enumerate}

\section{Five-dimensional bulk instabilities}

Let us consider the $F(R)$-gravity theory in five dimensions;
\begin{equation}
\label{A}
S=\frac{1}{2\kappa_{5}^{2}}\int \sqrt{-g}\left[ F^{(5)}(R)+S_{m} \right] \, , 
\end{equation} 
where $\kappa_{5}$ is the five-dimensional gravitational constant and $S_{m}$ is the action of the 
matter.
The equations of motion in the vacuum are given by 
\begin{equation} 
\label{AA}
F^{(5)}_{R}(R)
\left(R_{\mu\nu}-\frac{1}{2}Rg_{\mu\nu} \right) 
=\frac{1}{2}g_{\mu\nu} \left[ F^{(5)}(R)-RF_{R}^{(5)}(R) \right]
+\left[ \nabla_{\mu}\nabla_{\nu}-g_{\mu\nu}\Box \right]F_{R}^{(5)}(R) \, ,
\end{equation}
where $F^{(5)}_{R}= d F^{(5)}/ d R$.
Especially if we assume that the metric is covariantly constant, that is, 
$R_{\mu\nu} \propto K g_{\mu\nu}$ with a constant $K$, we find 
\be 
\label{AA2} 
0= RF^{(5)}_R (R) - \frac{5}{2}F^{(5)} (R)\, .
\ee
We denote the solution of Eq.~(\ref{AA2}) as $R=R_0$ and define the length parameter $l$ by 
$R_{0}=20/l^{2}$.
We should note that the metric of the Schwarzschild-de Sitter solution is covariantly constant 
and given by, 
\be 
\label{B} 
ds_\mathrm{SdS, (5)}^{2}= \frac{1}{h(a)}da^{2}-h(a)dt^{2}+a^{2}
d\Omega_{(3)}^{2}\, ,\quad
h(a)= 1-\frac{a^{2}}{l^{2}}-\frac{16\pi G_{(5)} M}{3 a^{2}} \, .
\ee
Here $M$ corresponds to the mass of the black hole and 
$G_{(5)}$ is defined by $8\pi G_{(5)} = \kappa_{5}^2$.
The space-time expressed by the metric (\ref{B}) has two horizons at 
\be 
\label{B1}
a^2 = a_\pm ^2 = \frac{l^2}{2}
\left\{ 1 \pm \sqrt{ 1 - \frac{64\pi G_{(5)} M}{3 l^2}} \right\} \, .
\ee
The two horizons degenerate in the limit, 
\be 
\label{B2} \frac{64\pi G_{(5)} M}{3 l^2} \to 1\, , 
\ee 
and we obtain the degenerate Schwarzschild-de Sitter (Nariai) solution.
The metric in the Nariai space-time is given by 
\begin{equation} 
\label{dsdmL} 
ds^{2} =\frac{1}{\Lambda}\left(-\sin^{2}\chi d\psi^{2}+d\chi^{2}
+d\Omega_{(3)}^2 \right) \, ,
\end{equation}
where there are the horizons at $\chi=0,\, \pi$ and $\Lambda=\frac{2}{l^2}$.
Let us perform the coordinate tranformation $\chi= \arccos \zeta$, 
\be 
\label{B3} 
ds^{2}=-\frac{1}{\Lambda} \left(1-\zeta^{2} \right)d\psi^{2}
+\frac{d\zeta^{2}}{\Lambda \left(1-\zeta^{2} \right)}
+\frac{1}{\Lambda}d\Omega_{(3)}^2 \, ,
\ee
which is singular at $\zeta=\pm 1$.
By changing the coordinate $\zeta=\tanh \xi$, the metric can be rewritten as, 
\be 
\label{B4} 
ds^{2}=\frac{1}{\Lambda \cosh^2 \xi} \left( - d \psi^2 + d\xi^2 \right)
+\frac{1}{\Lambda}d\Omega_{(3)}^2 \, .
\ee
We often analytically continue the coordinates by 
\be 
\label{B5} 
\psi = ix\, , \quad \zeta = i\tau\, , 
\ee 
and we obtain the following metric 
\begin{equation} 
\label{dsNa} 
ds^{2}=-\frac{1}{\Lambda \cos^{2}\tau} \left(-d\tau^{2}+dx^{2} \right) 
+\frac{1}{\Lambda}d\Omega_{(3)}^2 \, .
\end{equation}
Of course, after the analytic continuation, the obtained space is a solution of the equations 
although the topology is changed.
This expression of the metric was used in \cite{Bousso:1997wi}.

In order to consider the perturbation, we now consider the general metric in the following form, 
\begin{equation} 
\label{dse} 
ds^{2}=\e^{2\rho(x,\tau)} \left( -d\tau^{2}+dx^{2} \right)
+\e^{-2\phi(x,\tau)}d\Omega_{(3)}^2 \, ,
\end{equation}
which generalizes the Nariai's metric in Eq.(\ref{dsNa})
with generic functions $\rho(x,\tau), \phi(x,\tau)$.

Then the equation of motion can be decomposed in components as 
\begin{align} 
\label{e1} 
0=&-\frac{\e^{2\rho}}{2}F^{(5)} - \left( -\ddot{\rho}
+3\ddot{\phi}+\rho''-3\dot{\phi}^{2}-3\dot{\rho}\dot{\phi}-3\rho'\phi'
\right)F^{(5)}_{R}
+\ddot{F}^{(5)}_{R} \nn
& -\dot{\rho}\dot{F}^{(5)}_{R}-\rho' \left(F^{(5)}_{R} \right)'
+\e^{2\phi}\left[-\frac{\partial}{\partial \tau}\left(\e^{-2\phi}\dot{F}^{(5)}_{R}
\right)+\left(\e^{-2\phi}(F^{(5)}_{R})' \right)' \right] \, , \\ 
\label{e2}
0 =&\frac{\e^{2\rho}}{2}F^{(5)}-\left(-\rho''+3\phi''+\ddot{\rho}-3\phi'^{2}
 -3\rho'\phi'-3\dot{\rho}\dot{\phi}\right)F^{(5)}_{R}+{F^{(5)}_{R}}'' \nn 
 &-\dot{\rho}\dot{F}^{(5)}_{R}-\rho' \left(F^{(5)}_{R} \right)'
  -\e^{2\phi}\left[-\frac{\partial}{\partial
\tau}\left(\e^{-2\phi}\dot{F}^{(5)}_{R}\right)
+\left(\e^{-2\phi} \left(F^{(5)}_{R} \right)'\right)' \right] \, , \\
\label{e3}
0=&- \left(3\dot{\phi}'-3\phi'\dot{\phi}-3\rho'\dot{\phi}-3\dot{\rho}\phi' 
\right)F^{(5)}_{R}
+\frac{\partial^{2}F^{(5)}_{R}}{\partial x \partial \tau}
 -\dot{\rho} \left(F^{(5)}_{R} \right)'-\rho'\dot{F}^{(5)}_{R} \, , \\ 
 \label{e4} 0=&\frac{\e^{-2\phi}}{2}F^{(5)}-\e^{-2(\rho+\phi)}
\left( -\ddot{\phi}+\phi''+3\dot{\phi}^{2}-3\phi'^{2}
\right)F^{(5)}_{R} -F^{(5)}_{R}
+\e^{-2(\rho+\phi)}\left(\dot{\phi}\dot{F}^{(5)}_{R}
  -\phi'{F^{(5)}_{RR}}' \right) \nn
& -\e^{-2\rho}\left[-\frac{\partial}{\partial
\tau}\left(\e^{-2\phi}\dot{F}^{(5)}_{R}\right)
+\left(\e^{-2\phi}{F^{(5)}_{RR}}' \right)' \right]\, ,
\end{align}
where
$F'= \frac{\partial F}{\partial x}$ and $\dot{F}=\frac{\partial F}{\partial \tau}$ 
and we have used the expressions of the curvatures (\ref{Ncurvature}) in the Appendix \ref{A1}.

We consider the perturbations at the first order around the Nariai background 
Eq.(\ref{dsNa}) with 
$R_{0}=\frac{20}{l^2}$, 
\begin{align} 
\label{p1} 
0=& \frac{-F^{(5)}_{R}(R_{0}) +2\Lambda F^{(5)}_{RR}(R_{0})}{2\Lambda \cos^{2}\tau}\delta R
  -\frac{F^{(5)}(R_{0})}{\Lambda \cos^{2}\tau}\delta \rho
  -F^{(5)}_{R}(R_{0}) \left(-\delta \ddot{\rho}+3\delta \ddot{\phi}
+\delta \rho'' -3\tan \tau \delta \dot{\phi}\right) \nn
& -\tan \tau F^{(5)}_{RR}(R_{0})\delta \dot{R}+F^{(5)}_{RR}(R_{0})\delta R'' \, , \\ 
\label{p2} 
0=& -\frac{-F^{(5)}_{R}(R_{0})
+2\Lambda F^{(5)}_{RR}(R_{0})}{2\Lambda \cos^{2}\tau}\delta R 
+\frac{F^{(5)}(R_{0})}{\Lambda \cos^{2}\tau}\delta \rho
  -F^{(5)}_{R}(R_{0})
\left(\delta \ddot{\rho}+3\delta \phi''-\delta \rho''
  -3\tan \tau \delta \dot{\phi} \right) \nn 
& -\tan \tau F^{(5)}_{RR}(R_{0})\delta \dot{R}+F^{(5)}_{RR}(R_{0})\delta R''\, , \\ 
\label{p3} 
0=& -3F^{(5)}_{R}(R_{0}) \left(\delta \dot{\phi}'-\tan \tau \delta \phi' \right)
+F^{(5)}_{RR}(R_{0}) \left(\delta \dot{R}'-\tan \tau \delta R' \right)
\, , \\
\label{p4}
0=& - \frac{-F^{(5)}_{R}(R_{0})
+2\Lambda F^{(5)}_{RR}(R_{0})}{2\Lambda \cos^{2}\tau }\delta R
 -\frac{F^{(5)}(R_{0})}{\Lambda \cos^{2}\tau }\delta \phi
 -F^{(5)}_{R}(R_{0})\left(-\delta \ddot{\phi}+\delta \phi'' \right) \nn
& - F^{(5)}_{RR}(R_{0})(-\delta \ddot{R}+\delta R'') \, .
\end{align}
The perturbation of the scalar curvature $\delta R$ is given in terms of $\delta \rho$ and 
$\delta \phi$ as follows, 
\begin{equation} 
\label{dR} 
\delta R=4\Lambda(-\delta \rho+\delta \phi)
+\Lambda \cos^{2}\tau(2\delta \ddot{\rho}-2\delta \rho''-6\delta \ddot{\phi}+6\delta \phi'')\,.
\end{equation}
Therefore the four equations (\ref{p1}), (\ref{p2}), (\ref{p3}), and (\ref{p4}) include only two 
$\delta \phi$ and $\delta \rho$, which tell that only two equations in the four equations 
(\ref{p1}), (\ref{p2}), (\ref{p3}), and (\ref{p4}) should be independent ones.

One can find that Eq.~(\ref{p3}) can be easily integrated 
\begin{equation} 
\label{C1} 
\delta R=3\frac{F^{(5)}_{R}(R_{0})}{F^{(5)}_{RR}(R_{0})}\delta \phi
+\frac{c_{1}(x)}{\cos \tau}+c_{2}(\tau)\, .
\end{equation}
Here $c_1(x)$ and $c_2(\tau)$ are arbitrary functions but because $\delta R$ should vanish 
when both of $\delta \rho$ and $\delta \phi$ vanish as seen from (\ref{dR}), we can put 
$c_1(x)=c_2(\tau)=0$.

Then, one can directly consider Eq.(\ref{dR}):
Substituting in it $\delta R(\delta \phi)$ obtained in Eq.(\ref{C1}), we find a simple equation 
\begin{equation} 
\label{E} 
\left(\Box+\frac{M^{2}}{\cos^{2}\tau}\right)\delta \phi=0 \, , \quad 
\Box \equiv - \frac{\partial^2}{\partial \tau^2} + \frac{\partial^2}{\partial x^2}\, .
\end{equation}
Here
\begin{equation}
\label{alpha}
M^{2}=\frac{1}{2}\frac{4\alpha-1}{\alpha}\, , \quad \alpha 
=\frac{4\Lambda F^{(5)}_{RR}(R_{0})}{F^{(5)}_{R} \left(R_{0} \right)} 
=\frac{F(R_{0})F_{RR}(R_{0})}{[F_{R}(R_{0})]^{2}} \, .
\end{equation}
Eq.~(\ref{E}) is nothing but a time-dependent Klein-Gordon equation for the $\delta\phi$ mode, 
with an effective oscillating mass term in time.
An explicit solution of (\ref{E}) is given by 
\be 
\label{S1} 
\delta \phi = \phi_0 \cos \left( \beta x \right) \cos^\beta \tau \, .
\ee
Here $\beta$ is given by solving the equation $M^2 = \beta \left( \beta - 1 \right)$.
The anti-evaporation corresponds to the increasing of the radius of the apparent horizon, which is 
defined by the condition, 
\begin{equation} 
\label{condition} 
\nabla \delta \phi \cdot \nabla \delta \phi=0 \, .
\end{equation}
In other words, it is imposed that the (flat) gradient of the two-sphere size is null.
By using the solution in (\ref{S1}), we find $\tan \beta x = \tan \tau$, that is, $\beta x = \tau$.
Therefore on the apparent horizon, we find 
\be 
\label{S2} 
\delta \phi = \phi_0 \cos^{\beta+1} \tau \, .
\ee
Because the horizon radius $r_H$ is given by $r_H = \e^{- \phi}$, we find 
\be 
\label{S3} 
r_H = \frac{\e^{- \phi_0 \cos^{\beta+1} \tau}}{\sqrt{\Lambda}}\, .
\ee
Then if $\beta < -1$, the horizon grows up, which corresponds to the anti-evaporation depending 
on the sign of $\phi_{0}$.
The sign could be determined by the initial condition of the perturbation.
On the other hand, it is also possible the case in which $\beta,\omega$ are complex parameters.
In this case, solutions of perturbed equations read 
\be 
\label{NS} 
\delta \phi={\rm Re}\left\{(C_{1}\e^{\beta t}+C_{2}\e^{-\beta t})\e^{\beta x} \right\}\, , 
\ee 
where $C_{1,2}$ are complex numbers.
$\delta \phi$ always increase in time for $C_{1}\neq 0$ because of
${\rm Re}\beta>0$.
This means that the Nariai solution is unstable also in this region of parameters.
A particular class among possible complex parameter solutions is 
\be 
\label{NS2} 
\delta \phi= \phi_{0}\left\{\e^{\frac{-t+x}{2}}\left(\cos \frac{\gamma(t-x)}{2}
+\frac{1}{\gamma}\sin \frac{\gamma(t-x)}{2}\right) 
+\e^{\frac{t+x}{2}}\left(\cos \frac{\gamma(t+x)}{2}
 -\frac{1}{\gamma}\sin \frac{\gamma(t+x)}{2}\right)\right\}\,,
\ee
where 
$\beta\equiv \frac{1}{2}(1+i\gamma)$ and $\gamma\equiv \pm \sqrt{\frac{2-9\alpha}{\alpha}}$.

On the horizon, the fluctuations must satisfy the condition 
$\frac{\phi_{0}^{2}}{2}\gamma^{2}\e^{x}\sin\frac{\gamma(t-x)}{2}
\sin \frac{\gamma(t+x)}{2}=0$,
which corresponds to two classes of solutions with $x=\mp t+\frac{2n\pi}{\gamma}$, 
\be 
\label{NS2} 
\delta \phi=\phi_{0}(-1)^{n}\left\{ \e^{\frac{n\pi}{\gamma}}
+\e^{\mp t+\frac{n\pi}{\gamma}}\left(\cos\gamma t\mp
\frac{1}{\gamma}\sin \gamma t \right)\right\}\,, 
\ee 
which implies an oscillating horizon radius.

Let us consider a class of $F^{(5)}(R)$ models 
\begin{equation} 
\label{fRc} 
F^{(5)}(R)=\frac{R}{2\kappa^{2}}+f_{2}R^{2}+f_{0}\mathcal{M}^{5-2n}R^{n}\, .
\end{equation}
Here $f_2$ and $\mathcal{M}$ are constants with a mass dimension and $f_0$ is a dimensionless 
constant.
In this case, $\alpha$ is given by
\begin{equation}
\label{ps}
\alpha=\frac{4\Lambda
\left( 2f_{2}+n(n-1)f_{0}\mathcal{M}^{5-2n}R_{0}^{n-2} \right)}{1/2\kappa^{2}+2f_{2}R_{0} 
+nf_{0}M^{5-2n}R_{0}^{n-1}}\, .
\end{equation}
Then $\beta$ is given by
\be
\label{beta1}
\beta^2 - \beta =\frac{1}{2\alpha}(4\alpha-1) \, , \ee that is \be \label{beta2} \beta_{\pm} 
= \frac{1}{2} \left( 1 \pm \sqrt{ \frac{9\alpha-2}{\alpha}} \right) \, .
\ee
Then the condition of the anti-evaporation $\beta<-1$ (for $\phi_{0}<0$) can be satisfied only by 
$\beta_{-}$ and for $\alpha<0$.
On the other hand, for $\beta$ as a complex parameter in Eq.(\ref{NS2}), the oscillation instabilities 
are obtained for $0<\alpha<2/9$.
In this case, evaporation and antievaporation phases are iterated.

\subsection{Brane dynamics in the bulk }

We now consider the $F^{(d+1)}(R)$ gravity in the $d+1$ dimensional space-time $M$ with 
$d$ dimensional boundary $B$, whose action is given by 
\begin{equation} 
\label{FRd1} 
S=\frac{1}{2\kappa^2}\int_M d^{d+1} x \sqrt{-g} F^{(d+1)}(R)\, , 
\end{equation} 
which can be rewritten in the scalar-tensor form.
We begin by rewriting the action (\ref{FRd1}) by introducing the auxiliary field $A$ as follows, 
\begin{equation} 
\label{FRd2} 
S=\frac{1}{2\kappa^2}\int d^{d+1} x \sqrt{-g}
\left\{{F^{(d+1)}}'(A)\left(R-A\right) + F^{(d+1)}(A)\right\}\, .
\end{equation}
By the variation of the action with respect to $A$, we obtain the equation $A=R$ and by 
substituting the obtained expression $A=R$ into the action (\ref{FRd2}), we find that the action in 
(\ref{FRd1}) is reproduced.
If we rescale the metric by conformal transformation, 
\begin{equation} 
\label{JGRG22} 
g_{\mu\nu}\to \e^\sigma g_{\mu\nu}\, ,\quad \sigma = -\ln {F^{(d+1)}}'(A)\, , 
\end{equation} 
we obtain the action in the Einstein frame, 
\begin{align} 
\label{FRd3} 
S_E =& \frac{1}{2\kappa^2}\int_M d^{d+1} x \sqrt{-g} \left(R
 - (d-1)\Box \sigma - \frac{(d-2)(d-1)}{4}\partial^\mu \sigma \partial_\mu \sigma
 - V(\sigma)\right) \nn 
= & \frac{1}{2\kappa^2}\int_M d^{d+1} x \sqrt{-g} \left(R 
 - \frac{(d-2)(d-1)}{4}\partial^\mu \sigma \partial_\mu \sigma
 - V(\sigma)\right)
+ (d-1) \int_B d^d x \sqrt{-\hat g} n^\mu \partial_\mu \sigma \, , \nn
V(\sigma) =& \e^\sigma g\left(\e^{-\sigma}\right)
 - \e^{2\sigma} f\left(g\left(\e^{-\sigma}\right)\right)
 = \frac{A}{{F^{(d+1)}}'(A)} - \frac{F^{(d+1)}(A)}{{F^{(d+1)}}'(A)^2}\, .
\end{align}
Here $g\left(\e^{-\sigma}\right)$ is given by solving the equation 
$\sigma = - \ln {F^{(d+1)}}'(A)$ as $A=g\left(\e^{-\sigma}\right)$.
By the integration of the term $\Box \sigma$, there appears the boundary term, where $n^\mu$ is 
the unit vector perpendicular to the boundary and the direction of the vector is inside.
Furthermore $\hat g_{\mu\nu}$ is the metric induced on the boundary, 
$\hat g_{\mu\nu} = g_{\mu\nu} - n_\mu n_\nu$.
The existence of the boundary makes the variational principle with respect to $\sigma$ ill-defined, 
we cancel the term by introducing the boundary action 
\begin{equation} 
\label{FRd4} 
S_B = - (d-1) \int_B d^d x \sqrt{-\hat g} n^\mu \partial_\mu \sigma \, .
\end{equation}
Then one may forget the boundary term,
\begin{equation}
\label{FRd5}
S_E \to S_E + S_B = \frac{1}{2\kappa^2}\int_M d^{d+1} x \sqrt{-g} 
\left(R - \frac{(d-2)(d-1)}{4}\partial^\mu \sigma \partial_\mu \sigma
 - V(\sigma)\right) \, .
\end{equation}
As is well-known, because the scalar curvature $R$ includes the second derivative term, the 
variational principle is still ill-defined in the space-time with boundary \cite{Gibbons:1976ue} 
(see also, Refs.~\cite{Chakraborty:2014xla,Chakraborty:2015bja,Nojiri:2004bx,Deruelle:2007pt}).

Because the variation of the scalar curvature with respect to the metric is given by 
\begin{equation} 
\label{EE1} 
R = -\delta g_{\mu\nu} R^{\mu\nu}
+ g^{\sigma\nu} \left(\nabla_\mu \delta{\Gamma}^\mu_{\sigma\nu}
 - \nabla_\sigma \delta \Gamma^\mu_{\mu\nu} \right) \, , 
\end{equation} 
the variation of the action with respect to the metric is given by, 
\begin{equation} 
\label{EE2} 
\delta S_\mathrm{E}= \frac{1}{2\kappa^2} \int d^{d+1} x \sqrt{-g} Q^{\mu\nu} \delta g_{\mu\nu}
+ \frac{1}{2\kappa^2} \int_B d^d x \sqrt{-\hat g}
g^{\sigma\nu} \left( - n_\mu \delta \Gamma ^\mu_{\sigma\nu}
+ n_\sigma \delta \Gamma ^\mu_{\mu\nu} \right)\, .
\end{equation}
Here the Einstein equation in the bulk is given by $Q_{\mu\nu}=0$.
Then the variational principle becomes well-defined if we add the following boundary term, 
\begin{equation} 
\label{E3} 
\tilde S_b=- \frac{1}{2\kappa^2} \int_B d^d x \sqrt{-\hat g} g^{\sigma\nu} \left(
 - n_\mu \Gamma ^\mu_{\sigma\nu}+ n_\sigma \Gamma ^\mu_{\mu\nu} \right)\, .
\end{equation}
Although the above boundary term (\ref{E3}) is not invariant under the reparametrization, because 
\begin{equation} 
\label{E4} 
\nabla_\mu n_\nu=\partial_\mu n_\nu
 - \Gamma_{\mu\nu}^\lambda n_\lambda \ ,\quad \nabla_\mu n^\nu=\partial_\mu n^\nu
+ \Gamma_{\mu\lambda}^\nu n^\lambda\, ,
\end{equation}
we find
\begin{equation}
\label{E5}
g^{\sigma\nu} \left( - n_\mu \Gamma ^\mu_{\sigma\nu}
+ n_\sigma \Gamma ^\mu_{\mu\nu} \right)
= - \partial_\mu n^\mu - 2g^{\delta\rho}\partial_\delta n_\rho
\nabla_\mu n^\mu\, ,
\end{equation}
which is just equal to $\nabla_{\mu}n^{\mu}$ on the boundary  
\cite{Chakraborty:2014xla,Chakraborty:2015bja,Nojiri:2004bx,Deruelle:2007pt,Gibbons:1976ue} . 
Therefore we can replace the boundary term (\ref{E3}) by the Gibbons-Hawking boundary term, 
\begin{equation} 
\label{E6} 
S_\mathrm{GH} = \frac{1}{\kappa^2} \int_B d^d x \sqrt{-\hat g}\nabla_\mu n^\mu\, .
\end{equation}

Let the boundary is defined by a function $f(x^\mu)$ as $f(x^\mu)=0$.
Then by the analogy of the relation between the electric field and the electric potential in the 
electromagnetism, we find that the vector $\left( \partial_\mu f\left( x^\mu \right) \right)$ is 
perpendicular to the boundary because 
$dx^\mu \partial_\mu f\left( x^\mu\right) = 0$ on the
boundary, which gives an expression for $n_\mu$ as 
\be 
\label{n1} 
n_\mu = \frac{ \partial_\mu f}{\sqrt{g^{\rho\sigma} \partial_\rho f \partial_\sigma f}}\, .
\ee
Then with respect to the variation of the metric, the variation of $n^\mu$ is given by 
\be 
\label{n2} 
\delta n_\mu = \frac{1}{2} \frac{ \partial_\mu f} {\left( g^{\rho\sigma} 
\partial_\rho f \partial_\sigma f\right)^{\frac{3}{2}}} \partial^\tau f \partial^\eta f 
\delta g_{\tau\eta} = \frac{1}{2} n_\mu n^\rho n^\sigma \delta g_{\rho\sigma}\, .
\ee
By using the expression in (\ref{n2}), one finds the variation of $\nabla_\mu n^\mu$ with respect to 
the metric, 
\be 
\label{n3} 
\delta \left( 2 \nabla_\mu n^\mu \right) = - 2\delta g_{\mu\nu} n^\mu n^\nu
 - n^\mu \nabla^\nu \delta g_{\mu\nu}
 - g^{\mu\nu} n_\rho \delta \Gamma^\rho_{\mu\nu}
+ n^\nu \delta \Gamma^\mu_{\mu\nu}\, .
\ee
The last two terms in (\ref{n3}) are necessary to make the variational principle well-defined but 
the second term $n^\mu \nabla^\nu \delta g_{\mu\nu}$ also may violate the variational principle.
By using the reparametrization invariance, however, we can choose the gauge condition so that 
$\nabla^\nu \delta g_{\mu\nu}=0$.

We may also add the following boundary term, 
\begin{equation} 
\label{EE3} 
S_\mathrm{BD} = \int_B d^d x \sqrt{-\hat g} \mathcal{L}_\mathrm{B} \, , 
\end{equation} 
The variation of the total action 
\begin{equation} 
\label{EE4} 
S_\mathrm{total} = S_\mathrm{E} + S_B + S_\mathrm{GH} + S_\mathrm{BD} \, , 
\end{equation} 
is given by 
\begin{equation} 
\label{EE5} 
\delta S_\mathrm{total} = \frac{1}{2\kappa^2} \int d^{d+1} x 
\sqrt{-g} Q^{\mu\nu} \delta g_{\mu\nu}+ \int_B d^d x \sqrt{-\hat g} \left[ \frac{1}{2\kappa^2}
\left( \frac{1}{2} \mathcal{K} \hat g^{\mu\nu} - \mathcal{K}^{\mu\nu} \right)
+ \frac{1}{2}T_\mathrm{B}^{\mu\nu} \right] \delta g_{\mu\nu}\, .
\end{equation}
Here we have defined the extrinsic curvature by 
$\mathcal{K}_{\mu\nu} \equiv \nabla_{\mu} n_{\nu}$ and 
$\mathcal{K} \equiv g^{\mu\nu} \mathcal{K}_{\mu\nu}$.
We also wrote the variation of $S_\mathrm{BD}$ as 
\begin{equation} 
\label{EE6} 
\delta S_\mathrm{BD} = \frac{1}{2}\int_B d^d x \sqrt{-\hat g} 
T_\mathrm{B}^{\mu\nu} \delta g_{\mu\nu}\, .
\end{equation}
Then on the boundary, we obtain the following equation, 
\begin{equation} 
\label{EE7}
0 = \frac{1}{2} \mathcal{K} \hat g^{\mu\nu} - \mathcal{K}^{\mu\nu}
+ \kappa^2 T_\mathrm{B}^{\mu\nu} \, ,
\end{equation}
which may be called the brane equation.
Especially if the boundary action $S_\mathrm{BD}$ consists of only the brane tension 
$\tilde\kappa$, 
\begin{equation} 
\label{EE8} 
S_\mathrm{B} = \frac{\tilde\kappa}{\kappa^2} \int_B d^d x \sqrt{-\hat g} \, , 
\end{equation} 
we find 
\begin{equation} 
\label{EE9}
0 = \frac{1}{2} \mathcal{K} \hat g^{\mu\nu} - \mathcal{K}^{\mu\nu}
+ \tilde\kappa g^{\mu\nu} \, ,
\end{equation}
which can be rewritten as,
\begin{equation}
\label{EE11}
0 = \frac{2}{d-2} \tilde\kappa \hat g^{\mu\nu} - \mathcal{K}^{\mu\nu} \, .
\end{equation}
If we consider the model which is given by gluing two space-time as 
in the Randall-Sundrum model \cite{Randall:1999ee,Randall:1999vf}, 
the contribution from the bulk doubles and therefore the Gibbons-Hawking term also doubles, 
\begin{equation} 
\label{EE12}
0 = \frac{2}{d-2} \tilde\kappa \hat g^{\mu\nu} - 2 \mathcal{K}^{\mu\nu} \, .
\end{equation}
Let us consider the following five-dimensional geometry, 
\begin{equation} 
\label{SAdS} 
ds^{2}_\mathrm{5}= g_{\mu\nu}dx^\mu dx^\nu = - \e^{2\rho} dt^2 + \e^{-2\rho} da^2
+ a^2 d \Omega_\mathrm{3}^2 \, .
\end{equation}
Here $d \Omega_\mathrm{3}^2 = \tilde g_{ij} dx^i dx^j$ expresses the metric of the unit sphere in 
two dimensions.
We now introduce a new time variable $\tau$ so that the following condition is satisfied, 
\begin{equation} 
\label{cd1}
 -\e^{2\rho}\left( \frac{\partial t}{\partial \tau} \right)^2
+\e^{-2\rho}\left( \frac{\partial a}{\partial \tau} \right)^2 = -1 \ .
\end{equation}
Then we obtain the following FRW metric
\begin{equation}
\label{met1}
ds^{2}_4=\tilde g_\mathrm{ij} dx^i dx^j = -d\tau ^2 +a^2 d \Omega_\mathrm{3}^2 \, .
\end{equation}
Then
\begin{equation}
\label{nmu}
n^\mu =\left( - \e^{-2\rho} \frac{\partial a}{\partial \tau},
 -\e^{2\rho} \frac{\partial t}{\partial\tau}, 0,0,0 \right)\, .
\end{equation}
Because
\begin{equation}
\label{dn}
\mathcal{K}_{ij} = \frac{\kappa}{2}\e^{4\rho} a \tilde g_{ij} \frac{dt}{d\tau}\, , 
\end{equation} 
from Eq.~(\ref{EE9}), we obtain, 
\begin{equation} 
\label{dn2} 
\e^{2\rho} \frac{dt}{d\tau} = - \frac{\tilde\kappa}{2} a \, .
\end{equation}
Using (\ref{cd1}) and defining the Hubble rate by $H= \frac{1}{a} \frac{da}{d\tau}$, one finds the 
following FRW equation for the brane, 
\begin{equation} 
\label{HH0}
H^2 = - \frac{\e^{2\rho(a)}}{a^2} + \frac{\tilde\kappa^2}{4}\, .
\end{equation}
Then in case of the Schwarzschild-de Sitter black hole, 
\begin{equation} 
\label{SdS} 
\e^{2\rho}= \frac{1}{a^{2} }\left( -\mu + a^{2} - \frac{a^4}{l_\mathrm{dS}^2} \right) \, , 
\end{equation} 
we obtain 
\begin{equation} 
\label{H4} H^2= \frac{1}{ l_\mathrm{dS}^2} - \frac{1}{a^2} + \frac{\mu}{a^4}
+ \frac{\kappa^2}{4} \, .
\end{equation}
Here $l_\mathrm{dS}$ is the curvature radius of the de Sitter space-time and $\mu$ is the black 
hole mass.
On the other hand, in the Schwarzschild-AdS black hole, 
\begin{equation} 
\label{AdSS} 
\e^{2\rho}= \frac{1}{a^{2} }\left( -\mu + a^{2} + \frac{a^4}{l_\mathrm{AdS}^2} \right) \, .
\end{equation}
we obtain,
\begin{align}
\label{F01}
H^2= - \frac{1}{l_\mathrm{AdS}^2} - \frac{1}{a^2}
+ \frac{\mu}{a^4} + \frac{\kappa^2}{4}\, .
\end{align}
In the Jordan frame, the metiric is given by 
\begin{equation} 
\label{dsJ} 
ds^2_{\mathrm{J}4} = {F^{(5)}}'(R) ds^{2}_4=\left( -d\tau ^2 +a^2 d
\Omega_\mathrm{3}^2 \right)\, .
\end{equation}
Because the scalar curvature is a constant in the Schwarzschild-(anti-)de Sitter space-time, 
${F^{(5)}}'(R)$ can be absorbed into the redefinition of $\tau$ and $a$, 
\begin{equation} 
\label{dsJ2} 
d\tilde\tau \equiv dt \sqrt{{F^{(5)}}'(R)} \, , \quad \tilde a \equiv a \sqrt{{F^{(5)}}'(R)} \, .
\end{equation}
Then the qualitative properties are not changed in the Jordan frame compared with the Einstein 
frame.
We should also note that the motion of the brane does not depend on the detailed structure of 
$F^{(5)}(R)$.

In the Nariai space, the radius $a$ is a constant and therefore $H=0$.
Furthermore in the Nariai space, we find $\e^{2\rho(a)}=0$ and therefore
Eq.~(\ref{HH0}) shows that the brane tension $\tilde\kappa$ should vanish.
That is, if and only if the tension vanished, the brane can exist.
The non-vanishing tension might be cancelled with the contribution from the trace anomaly by 
tuning the brane tension.
We should note, however, that there should not be any (FRW) dynamics of the brane in the Nariai 
space.

However, the anti-evaporation may induce the dynamics of the brane.
For the metric (\ref{dse}), one gets the expressions of the connection in (\ref{Nconnections}).
We introduce a new time coordinate $\tilde t$ in the metric (\ref{dse}) as follows, 
\be 
\label{NFRW1} 
d\tilde t^2 \equiv \e^{2\rho} \left( d\tau^2 - dx^2 \right) \, .
\ee
Then the metric(\ref{dse}) reduces to the form of the FRW-like metric, 
\begin{equation} 
\label{NFRW2} 
ds^{2}= - d\tau^2 +\e^{-2\phi(x,\tau)}d\Omega_{(3)}^2 \, , 
\end{equation} 
if we identify $\e^{-\phi(x,\tau)}$ with the scale factor $a$, $a=\e^{-\phi(x,\tau)}$.
Then the unit vector perpendicular to the brane is given by 
\be 
\label{NFRW3} 
n^\mu =\left( - \e^{-2\rho} \frac{\partial x}{\partial \tilde t},
  -\e^{-2\rho} \frac{\partial \tau}{\partial \tilde t}, 0,0,0 \right)\, ,
\ee
and the $(i,j)$ $\left(i,j=1,2,3\right)$ components Eq.~(\ref{EE9}) give 
\be 
\label{NFRW4}
 - \e^{-2\rho} \frac{\partial \phi}{\partial \tau} \frac{\partial x}{\partial \tilde t}
 - \e^{-2\rho} \frac{\partial \phi}{\partial x} \frac{\partial \tau}{\partial \tilde t} = \tilde \kappa \, .
\ee
As we discussed, in order that the brane exists in the Nariai space-time, we find 
$\tilde \kappa = 0$.
By using the solution in (\ref{S1}), and analytically recontinuing the coordinates $x\to - i\tau$, 
$\tau \to - ix$, if we assume 
\be 
\label{NFRW5} 
\phi = \ln \Lambda + \phi_0 \cosh \omega \tau \cosh^\beta x \, , 
\ee 
with $\omega^2 = \beta^2$, we find 
\be 
\label{NFRW6}
 - \omega \sinh \omega \tau \cosh^\beta x \frac{\partial x}{\partial \tilde t}
 - \beta \cosh \omega \tau \cosh^{\beta-1} x \sinh x \frac{\partial \tau}{\partial \tilde t} = 0\, , 
 \ee 
 that is, 
\be 
\label{NFRW7} 
\frac{\partial x}{\partial \tilde t} 
= - \frac{\beta \tanh x}{\omega \tanh \omega \tau} \frac{\partial \tau}{\partial \tilde t} \, .
\ee
Assuming that $x$ and $\tau$ only depend on $\tilde t$ on the brane, 
\be 
\label{NFRW8}
0 = \frac{1}{\beta \tanh x}\frac{dx}{d\tilde t}
+ \frac{1}{\omega \tanh \omega\tau} \frac{d\tau}{d\tilde t}
= \frac{d}{d\tilde t} \left( \frac{1}{\beta} \ln \sinh x
+ \ln \tanh \omega \tau \right)\, ,
\ee
that is $\frac{1}{\omega} \ln \sinh x+ \ln \sinh \omega \tau$ is a constant, which gives the 
trajectory of  the brane, 
\be 
\label{NFRW9} 
\sinh x = \frac{C}{\sinh^\beta \omega \tau}\, .
\ee
Here $C$ is a constant.
Of course, the expression in (\ref{NFRW9}) is valid as long as the perturbation 
$\delta\phi = \phi_0 \cosh \omega \tau \cosh^\beta x$ is small enough.
We should also note that because ${F^{(5)}}'(R)$ is not a constant due to the perturbation,
Eq.~(\ref{dsJ2}) also gives another source of the dynamics of the brane. However, that 
Eq.~(\ref{dsJ2}) gives only small correction to Eq.~(\ref{NFRW9}).

\section{Conclusion}

In this paper, we have studied the FRW brane lying in the degenerate Schwarzschild-de Sitter
(Nariai) black hole, in a five dimensional $F(R)$-gravity.
We have found that there cannot exist the brane in the Nariai bulk space except the case that the 
brane tension vanishes.
We have shown how the Nariai bulk is unstable in a large variety of $F(R)$-gravity models.
These instabilities
are divided in three classes; classical evaporation, classical anti-evaporation and infinite iterations 
of evaporations/anti-evaporations.
We have studied how these instabilities change the brane world-sheet dynamics.
%Finally, let us remark that the thermodynamical proprieties of these systems remain obscure.
%For example, an interpretation of classical evaporation and anti-evaporation from dS/CFT 
%correspondence is still missing.
%Perhaps, higher dimensional $F(R)$-gravity may suggest examples violating dS/CFT holographic 
%correspondence which deserves future investigation.

\begin{acknowledgments}

AA would like to thank University of Fudan for hospitality during the preparation of this letter.
This work  is partially supported  by the MIUR research grant ``Theoretical Astroparticle Physics" 
PRIN 2012CPPYP7 and by SdC Progetto speciale Multiasse ``La Societ\`a della Conoscenza" in 
Abruzzo PO FSE Abruzzo 2007-2013(AA), by MEXT KAKENHI Grant-in-Aid for Scientific Research 
on Innovative Areas ``Cosmic Acceleration''  (No. 15H05890)(SN), by MINECO (Spain), project
FIS2013-44881 (SDO) and by the CSIC I-LINK1019 Project (SDO and SN).

\end{acknowledgments}

%\vspace{0.5cm}

\appendix

\section{Components of the Ricci tensors and Ricci scalar \label{A1}}

For the metric (\ref{dse}),
we write $\eta_{ab}=\mathrm{diag}(-1,0)$, $\left(a,b=\tau,x\right)$.
For the metric $d\Omega_{(3)}^2$ of three dimensional unit sphere, we also write as 
\be 
\label{3dS}
d\Omega_{(3)}^2 = \hat g_{ij} dx^i dx^j \, , \quad (i,j=1,2,3)\, .
\ee
Then we obtain $\hat R_{ij} = 2 \hat g_{ij}$.
Here $\hat R_{ij}$ is the Ricci curvature given by $\hat g_{ij}$.

Then we find the following expression of the connections, 
\be 
\label{Nconnections} 
\Gamma^a_{bc} = \delta^a_{\ b} \rho_{,c} + \delta^a_{\ c} \rho_{,b}
 - \eta_{bc} \rho^{,a}\, , \quad
\Gamma^a_{ij} = \e^{-2(\rho+\phi)} \hat g_{ij} \phi^{,a}\, , \quad 
\Gamma^i_{aj} = \Gamma^i_{ja} = - \delta^i_{\ j} \phi_{,a}\, , \quad 
\Gamma^i_{jk} = \hat \Gamma^i_{jk}\, .
\ee
Here $\hat \Gamma^i_{jk}$ is the connection given by $\hat g_{ij}$.
By using the expressions in (\ref{Nconnections}), the curvatures are given by 
\begin{align} 
\label{Ncurvature} 
R_{ab} =& 3 \phi_{,ab} - \eta_{ab} \Box \rho - 3 \left(\phi_{,a} \rho_{,b}
+ \phi_{,b} \rho_{,a} \right) + 3 \eta_{ab} \phi_{,c} \rho^{,c}
 - 3 \phi_{,a} \phi_{,b} \, , \nn
R_{ij} = & \hat R_{ij} + \hat g_{ij} \e^{-2 (\rho+\phi)} \left( \Box \phi
 - 3 \phi_{,a} \phi^{,a} \right)\, , \quad R_{ia}=R_{ai}=0\, , \nn R = & \e^{2\phi} \hat R
+ \e^{-2\rho} \left( 6 \Box \phi - 2 \Box \rho - 12 \phi_{,a} \phi^{,a} 
\right) \, .
\end{align}
We should note $\hat R=6$ because $\hat R_{ij} = 2 \hat g_{ij}$.

\end{document}